\documentclass{aa}
\usepackage{graphicx}
\usepackage{longtable}
\usepackage{lscape}

\def\gsim{\;\lower.6ex\hbox{$\sim$}\kern-7.75pt\raise.65ex\hbox{$>$}\;}
\def\lsim{\;\lower.6ex\hbox{$\sim$}\kern-7.75pt\raise.65ex\hbox{$<$}\;}
\def\teff{$T_{\rm eff}$}
\def\ebv{E$(B-V)$}

\begin{document}
\title{Na-O Anticorrelation and HB. II. The Na-O
anticorrelation in the globular cluster NGC 6752
\thanks{
Based on observations collected at ESO telescopes under programme 073.D-0211.
Tables 2, 3, 5 are only available in electronic form
at the CDS via anonymous ftp to cdsarc.u-strasbg.fr (130.79.128.5)
or via http://cdsweb.u-strasbg.fr/cgi-bin/qcat?J/A+A/}
 }

\author{
E. Carretta\inst{1},
A. Bragaglia\inst{1},
R.G. Gratton\inst{2},
S. Lucatello\inst{2}
\and
Y. Momany\inst{2}
}

\authorrunning{E. Carretta et al.}
\titlerunning{Na-O anticorrelation in NGC 6752}

\offprints{E. Carretta, eugenio.carretta@oabo.inaf.it}

\institute{
INAF - Osservatorio Astronomico di Bologna, Via Ranzani 1, I-40127
 Bologna, Italy
\and
INAF - Osservatorio Astronomico di Padova, Vicolo dell'Osservatorio 5, I-35122
 Padova, Italy
  }

\date{8 jan 2007}

\abstract{
We are studying the Na-O anticorrelation in several globular clusters of
different Horizontal Branch (HB) morphology in order to derive a possible
relation between (primordial) chemical inhomogeneities and morphological
parameters of the cluster population.
We used the multifiber spectrograph FLAMES on the ESO Very Large Telescope UT2
and derived atmospheric parameters and elemental abundances of Fe, O
and Na for about 150 red giant stars in the Galactic globular cluster NGC~6752.
The average metallicity we derive is [Fe/H]=$-$1.56, in agreement with
other results from red giants, but lower than obtained for dwarfs or early
subgiants.
In NGC 6752 there is not much space for an
intrinsic spread in metallicity: on average, the rms scatter in [Fe/H] is 
$0.037\pm 0.003$ dex, while the scatter expected on the basis of the major 
error sources is $0.039\pm 0.003$ dex.
The distribution of stars along the Na-O anticorrelation is different to what
was found in the first paper of this series for the globular cluster 
NGC 2808: in NGC 6752 it is skewed toward more Na-poor stars, and it 
resembles more the one in M 13.
Detailed modeling is required to clarify whether this difference may explain
the very different distributions of stars along the HB.
 }

\maketitle

\section{Introduction}

This is the second paper of a series aimed at uncovering and studying the
possible  existence of a second generation of stars in Galactic Globular
Clusters (GCs). As explained in the first paper of the series (Carretta et al.
2006, hereafter Paper I) dedicated to the unusual cluster NGC 2808, to reach
our goal we are  performing a systematic analysis of a large number of stars
(about 100 per cluster) in about 20 GCs, determining accurately and
homogeneously  abundances of Fe, Na and O.
The well known anticorrelation between the abundances of the two last light 
elements (see Kraft 1994
and Gratton et al. 2004 for reviews covering the early discoveries and
recent developments) is attributed to proton-capture reactions in the complete
CNO cycle (Ne-Na and Mg-Al chains). The fact that the anticorrelation
extends to unevolved stars, unable to mix internal nucleosynthetic products
to the surface (Gratton et al. 2001, Ramirez \& Cohen 2003, Cohen \& Melendez
2005, Carretta et al. 2005) has reinforced the primordial origin hypothesis 
for these
abundance anomalies. If a previous generation of stars has polluted material
from which the presently living stars formed, we may expect to see 
differences in the abundances for those elements involved in nuclear burning
in those pristine stars.   

D'Antona and Caloi (2004) have suggested that the distribution of stars
along the Na-O anticorrelation is related to the distribution of stars along
the HB, thus being a potential explanation of the second parameter effect. In
fact, He is produced in $p-$capture at high temperature, the mechanism
explaining the Na-O anticorrelation (Denisenkov and Denisenkova 1989; Langer et
al. 1993; Prantzos and Charbonnel 2006). Since stars with different He content
burn H at different rates, they are expected to have different main sequence
lifetimes. At a given age, stars of different masses will be climbing the RGB
and, if they loose mass at the same rate, they will end up in different
locations along the HB. D'Antona and Caloi (2004) studied a few
typical examples, supporting the original hypothesis.
In this series we wish to produce a large observational dataset to test this
scenario (and other possibilities) in depth.

Since we intend to derive the distribution function of the anticorrelation, not
simply the general shape, we require large and unbiased samples of stars in
each cluster. In particular, when selecting the targets we do not
try to enhance the possibility of including extreme cases in order to define the 
existence and shape of a (possible) anticorrelation. This
was instead what we did when we 
studied a few turnoff and subgiant branch stars in NGC~6397, 47~Tuc and
NGC~6752 itself (Gratton et al. 2001, Carretta et al. 2005), where
we selected stars
with likely strong or weak CN bands, using the Str\" omgren $c_1$ index.
A fiber instrument (like FLAMES) mounted at a large telescope (the
VLT) is ideal to reach our goal and the  
homogeneity of data acquisition, treatment and analysis is basic to
compare results for different clusters. In fact, our approach is to use the
same tools for abundance analysis (the same atomic parameters, the same
procedure to derive atmospheric parameters, the same package to reduce spectra 
and to measure equivalent widths, the same prescriptions for NLTE corrections,
the same set of solar reference abundances and so on) for a large number of
stars in many clusters. We stress that our main target is not
simply to uncover the Na-O anticorrelation in several galactic GCs: we
currently know that this is as ubiquitous phenomenon (see e.g.
Gratton et al. 2004 and Paper I). We intend
to use a homogeneous approach in order to get rid of features possibly arising 
from limited samples and/or not self-consistent analysis; we 
aim at selecting properties of the Na-O anticorrelation that might be linked
to global physical parameters of the GCs, once all the sample of objects
available to us will be analyzed.

In this paper we present our results on the Na-O anticorrelation among
Red Giant Branch (RGB) stars in  NGC 6752. This nearby, low-reddened
cluster has been widely studied in the past, although never with the present
goal in mind.
Its metallicity\footnote{We adopt the  usual spectroscopic notation, $i.e.$ 
[X]= log(X)$_{\rm star} -$ log(X)$_\odot$ for any abundance quantity X, and 
log $\epsilon$(X) = log (N$_{\rm X}$/N$_{\rm H}$) + 12.0 for absolute number
density abundances.} has been derived from high-resolution spectra by many
authors in the range [Fe/H] $\sim -1.4$ to $-1.6$ (see e.g., Pritzl et al.
2005 for a recent compilation).

There have also been several studies of the Na-O anticorrelation in NGC 6752
but they have always been based on (much) smaller samples than the one
presented here. For instance, 
Norris \& Da Costa (1995) compared Na, O abundances for six giants in this
cluster (plus three in NGC 6397, and in 47 Tuc) to the ones in $\omega$ Cen, and Yong et
al. (2003) studied 20 bright RGB stars. 
In particular, this is the $first$ GC for which a Na-O (and Mg-Al) 
anticorrelation has been found $also$ for unevolved and scarcely evolved stars.
Gratton et al. (2001) and Carretta et al. (2005) studied 18 stars near the main
sequence Turn-Off and on the subgiant branch, demonstrating that (at least part
of) the chemical inhomogeneities in GCs must be implanted in the stars and
cannot be explained by evolutionary processes. 
Deep (or extra) mixing can begin to act only when the star is on the
giant branch,  after the RGB bump, and cannot work for unevolved stars without
deep convective envelopes.
This result is also supported by  Grundahl et al. (2002), who presented
evidence of the Na-O and Mg-Al anticorrelations for giants below the RGB bump;
these objects were later reanalyzed by Yong et al. (2005), together with
the ones near the RGB tip, for a total of 38 RGB stars.
A summary of all the previously available information on the Na-O
anticorrelation in this cluster can be found in Carretta et al. (2005); that
paper deals mostly with  CNO abundances in unevolved stars and presents further
arguments in favour of a primordial origin for elemental variations, in
particular from an earlier generation of intermediate mass Asymptotic Giant
Branch stars (Ventura et al. 2001).

The present paper is organized as follows: an outline of the observations is
given in the next Section; the derivation of atmospheric  parameters and the
analysis are discussed in Sect. 3, whereas error estimates are given in Sect.
4.  Sect. 5 is devoted to the intrinsic scatter in Fe, the reddening, and the
results for the Na-O anticorrelation; a discussion is presented in Sect. 6
and a summary is given in Sect. 7.

\section{Observations and measures}

\subsection{Observations} 

Our data were collected (in Service mode)  with the ESO high resolution
multifiber spectrograph  FLAMES/GIRAFFE (Pasquini et al. 2002) mounted on VLT
UT2. Observations were done with two GIRAFFE setups, using the high-resolution
gratings HR11 (centered at 5728~\AA)  and HR13 (centered at 6273~\AA) to measure
the Na doublets at 5682-5688~\AA\ and 6154-6160~\AA, and the [O {\sc i}]
forbidden lines at 6300,  6363~\AA, respectively. The resolution is R=24200
(for HR11) and R=22500  (for HR13). We have one single exposure of 1750 seconds
for each grating.

\begin{table}
\caption{Log of the observations for NGC 6752. Date and time are UT, exposure
times are in seconds. For both exposures the field center is at
RA(2000)=19:10:51.780, Dec(2000)=$-$59:58:54.70}
\begin{tabular}{rcccc}
\hline
Grating &Date       &UT$_{beginning}$ &exptime &airmass \\
\hline
HR11   & 2004-06-25  & 09:54:29  & 1750 &1.787\\ 
HR13   & 2004-06-23  & 09:41:13  & 1750 &1.677\\ 
\hline
\end{tabular}

\label{t:log}
\end{table}

Our targets were selected among isolated RGB stars, 
using the photometry by Momany et al. (2004): we chose stars lying near the
RGB ridge line without any companion closer than 2\arcsec \ and brighter than
the star magnitude plus 2.
Not all the stars were observed with both gratings; on a grand
total of 151 different stars observed, we have 72 objects with spectra for both
gratings, 41  with only HR11 observations and 38 with only HR13
observations. 
Since the Na doublet at 6154-6160~\AA\ falls into the spectral range covered by
HR13, we could measure Na abundances for all target stars, whereas we could
expect to measure O abundances only  up to a maximum of 110 stars.

\begin{figure}
\centering
\includegraphics[bb=60 180 340 590, clip, scale=0.8]{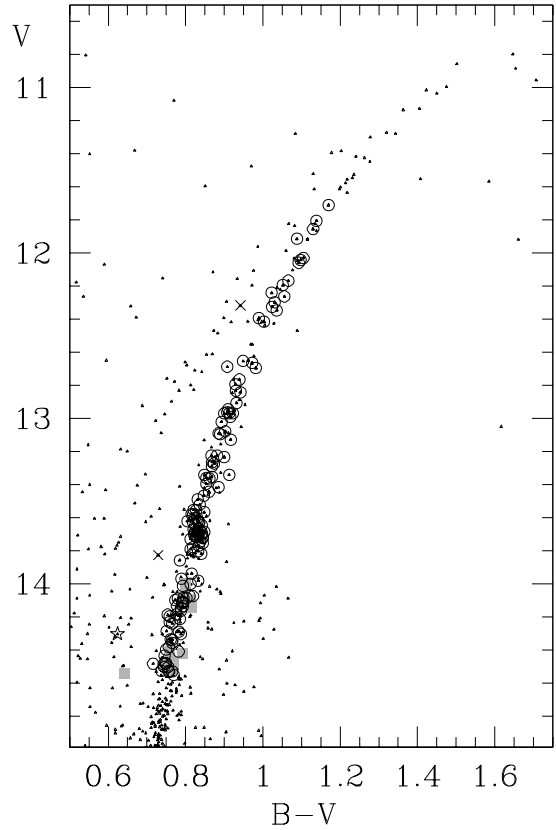}
\caption{$V,B-V$ CMD for NGC~6752 from Momany et al. (2004; dots);
observed stars are indicated by circles. Crosses indicate the two field stars, 
the open star symbol is for star 49370 whose spectrum shows double lines and
filled squares in grey tones indicate warm or faint objects with low $S/N$
spectra and no reliable abundance determinations.} 
\label{f:figcmd}
\end{figure}

Table~\ref{t:log} lists information about the two pointings, while a list of
all observed targets with coordinates, magnitudes and radial velocities (RVs)
is given in Table~\ref{t:coo} (the full table is available
only in electronic form). The $V,B-V$  colour magnitude
diagram (CMD) of our sample is shown in Figure~\ref{f:figcmd}; our targets
range from about $V$= 11.6 to 14.6, i.e. from about 1 magnitude below the RGB 
tip to about 1 magnitude below the RGB bump. Two field stars, accidentally
included in the sample, are indicated by
crosses, and the open star symbol is for a star (49370) whose spectrum presents
double lines (due to true binarity or to contamination from a nearby object).
These stars are disregarded from the following analysis.

Contamination from AGB stars is not of great concern in our sample. In
Figure~\ref{f:figcmd} the RGB and AGB sequences are clearly separated up to
$V,B-V\simeq(11.8,1.1)$ but very few stars of our sample are
located above this point in the CMD. 
Since AGB stars are expected to be about 10\% of RGB stars, contamination from 
interloper AGB stars should be minimal. This is confirmed,
a posteriori, by the very small scatter in the derived abundances, supporting
the soundness of the adopted parameters and the attribution of stars to the
first ascent red giant branch.

We used the 1-d, wavelength calibrated spectra as reduced by the dedicated
GIRAFFE pipeline (BLDRS  v0.5.3, written at the Geneva Observatory, see
{\em http://girbldrs.sourceforge.net}). Radial velocities were 
measured using the GIRAFFE pipeline, which performs cross correlations 
of the observed spectra with artificial spectral templates.
Further analysis was done in {\sc IRAF}\footnote{
IRAF is distributed by the National Optical Astronomical
Observatory, which are operated by the Association of Universities for
Research in Astronomy, under contract with the National Science
Foundation }: we subtracted the background using the 10 fibers dedicated to the
sky, rectified the spectra and shifted them to zero RV. 
Before this final step, we corrected the HR13 spectra for contamination from 
telluric features using a synthetic spectrum adapted to our resolution
and the intensity of telluric absorption, as we did in Paper I. 

There is a very small systematic difference between RVs in HR11 and HR13  (on
average about 0.3 km~s$^{-1}$,  rms 0.05 km~s$^{-1}$), which however  has no
influence on our abundance analysis.  The cluster heliocentric average velocity,
computed eliminating only two obvious non members, is $-25.6$ (rms 6.2) \,km
s$^{-1}$, in good agreement with the value tabulated in the updated web catalog
of Harris (1996) and with the average velocity  $-23.8\pm 2.1$ \,km s$^{-1}$
found from seven stars observed with UVES by Gratton et al. (2005), who 
also discuss other literature measurements.

\begin{table*}
\caption{List and relevant information for the target stars observed in
NGC~6752.  ID, $B$, $V$ and coordinates (J2000) are taken from Momany et al.
(2004); $J$, $K$ are from the 2MASS catalog; radial velocities RV's (in km
s$^{-1}$) from both gratings are heliocentric; listed S/N values are per
pixel, computed from spectra acquired with HR11 when available, else with
HR13; stars with '*' in notes have $V-K$ colours that deviate from the ones
expected for RGB stars (see text). Stars with '$\dagger$' in notes have too
much few lines (being warm and/or of low $S/N$ to obtain  reliable measurements
and were dropped from further analysis. The complete Table is available
electronically; we show here a few lines for guidance.}
\begin{tabular}{rllccccrrrrr}
\hline
 Star         &RA (h m s)     & DEC (d p s)    &  V     &   B   &   J   &  K    &S/N  &RV(HR11) &RV(HR13)& HR    &Notes\\
\hline
  1217 & 19  12  0.740 & -60   0  11.85 & 13.556 &14.383 &11.735 &11.110 & 107 &-29.52 & -29.97  &  11,13 &	   \\
  1493 & 19  11 54.785 & -59  59  15.44 & 13.858 &14.643 &12.064 &11.465 & 115 &       & -27.22  &  13    &	 \\
  1584 & 19  11 52.324 & -59  58  52.85 & 14.241 &15.007 &12.491 &12.065 &  51 &-28.20 & -29.17  &  11,13 &	   \\
  1797 & 19  11 46.677 & -59  58   8.83 & 13.791 &14.604 &12.003 &11.442 & 147 &       & -27.43  &  13    &	 \\
  2097 & 19  11 46.252 & -59  56  58.97 & 12.241 &13.264 &10.108 & 9.390 & 253 &-22.21 & -22.59  &  11,13 &	   \\
  2162 & 19  11 43.044 & -59  56  45.39 & 13.343 &14.192 &11.433 &10.818 & 117 &-25.18 & -25.70  &  11,13 &	   \\
  2704 & 19  11 42.883 & -59  54  24.53 & 13.718 &14.545 &11.897 &11.301 &  81 &-25.94 & -26.07  &  11,13 &	   \\
  2817 & 19  11 53.262 & -59  53  50.04 & 13.939 &14.755 &12.166 &11.556 &  76 &-26.13 & -27.50  &  11,13 &	   \\
  3039 & 19  11 44.832 & -59  52  30.49 & 14.095 &14.909 &12.347 &11.788 &  41 &-29.44 &	&  11	 & $\dagger$ \\
  4602 & 19  10 41.235 & -60   4  36.66 & 13.778 &14.597 &11.969 &11.384 &  70 &-25.51 & -25.79  &  11,13 &	   \\
\hline
\end{tabular}
\label{t:coo}
\end{table*}

\subsection{Equivalent widths}

Equivalent widths ($EW$s) were measured as described in detail in 
Bragaglia et al. (2001), adopting a relationship between EW and FWHM as
described at length in that paper.
Particular care was devoted to the definition
of the local continuum around each line, a delicate task at the 
moderately limited
resolution of our spectra, especially for the coolest targets. The choice of
the continuum level is done by an iterative procedure that takes into account
only a given fraction of the possible points. 
After several checks we decided that
the optimal choice for NGC~6752 is represented by a fraction 
of 1 for stars warmer than 4850~K, of 0.7 for stars with $T_{\rm eff}$
between 4600 and 4850~K, and about 0.6 for stars cooler than 4600~K.
Note that these values well approximate the general relation that we adopt
throughout all this series of papers when analyzing GIRAFFE spectra.
A few stars (indicated in Table~\ref{t:coo}) had too few lines 
(these objects
are warm and/or have low S/N$\leq 60$) to obtain reliable measurements, hence we dropped
them from further analysis. Tables of measured $EW$s will be only available at
the CDS database.

\begin{figure}
\centering
\includegraphics[bb=55 196 400 690, clip, scale=0.55]{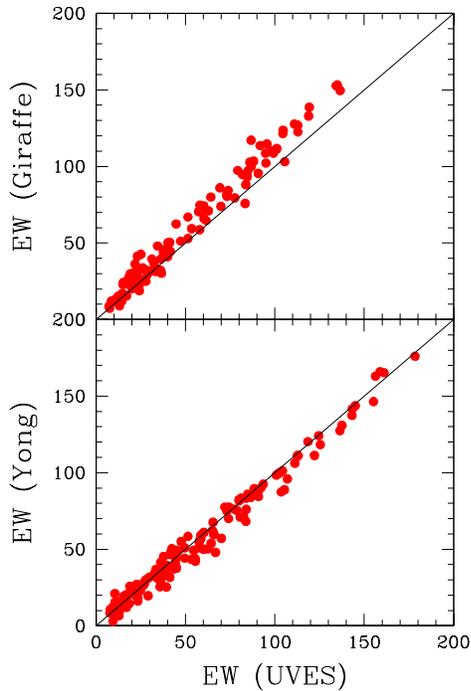}
\caption{Upper panel: comparison between the EWs derived from GIRAFFE and
UVES spectra for 8 stars in NGC~6752 observed in both modes. Lower panel:
comparison between EWs from UVES measured for two stars in common with Yong et
al. (2003).}
\label{f:confrontaew}
\end{figure}

To check the reliability of EWs measured on  the GIRAFFE spectra 
 we performed the following steps. While the
two GIRAFFE setups were obtained, we also observed 14 RGB stars in NGC~6752
through the dedicated fibers feeding the high resolution ($R\sim43000$)
spectrograph UVES. We employed the standard RED580 setup, obtaining
spectra in the range 4800-6800~\AA.\footnote{
The analysis of UVES spectra in NGC~6752 as well as in the
other globular clusters studied in the  present project will be presented 
separately.} 
We measured the 8 stars observed both with the GIRAFFE and the UVES
configuration with the above procedure; the comparison of the EWs is shown in
the upper panel of  Figure~\ref{f:confrontaew}. We found that on average EWs
from GIRAFFE are larger than those measured from UVES by
$+7.0\pm0.7$ m\AA\ (rms=7.2 m\AA\ from 114 lines). A linear regression
between the two sets of measurements gives
$EW_{\rm GIRAFFE}= 1.12 (\pm 0.02) \times EW_{\rm UVES}+1.17 (\pm 0.55)$ m\AA\
with
rms=5.89 m\AA\ and a correlation coefficient r=0.99. We inverted 
this relationship and used it to correct the $EW$s from GIRAFFE spectra to the system of
the higher resolution UVES spectra\footnote{
The difference between the UVES and
GIRAFFE $EW$s we found for NGC~6752 spectra is larger than observed for other
GCs analyzed in this series. This difference is not simply a function of
metallicity. Inspection of the original spectra showed that this difference is
not due to the measuring procedure and that lines in spectra of stars at the
same position in the CMD may be of different strength. We think that this is
due to small but significant effects intrinsic to the GIRAFFE spectra. Our
procedure reduces to a minimum the systematic error for each GC.}. 

In the lower panel of Figure~\ref{f:confrontaew} the $EW$s from our UVES
spectra are compared with those of two stars in common  with 
the sample by
Yong et al. (2003), taken with UVES at the highest possible spectrograph
resolution ($R=110000$) and with $S/N$ ranging from 250 to 150 per pixel.
The two stars are 11189, 23999
and mg24, mg15 in our catalogue and in Yong et al. (2003), respectively.
The agreement of our measures with their $EW$s, kindly provided by D. Yong (2006, private
communication) is excellent: on average the difference (in the sense us minus
Yong) is $+1.7 \pm 0.4$ m\AA\ with $\sigma=5.6$ m\AA\ from 157 lines.
This allows us to be quite confident about errors in the
continuum placement being small for the UVES spectra, at a negligible value, 
lower than 1\%\footnote{
$EW$s measured on moderate resolution spectra may be either 
overestimated due to contribution of blends, not recognized when compiling the
line list, or underestimated, because the nominal continuum used when
extracting EWs can be lower than the real value, due to veiling from weak
lines, not recognized as such due to the low resolution of the spectrum.}.

Thus, after correction to the UVES system, the $EW$s measured on  the GIRAFFE
spectra are not likely to be affected anymore by systematic effects due e.g., to
unaccounted blends possibly due to the moderate resolution.

In the following, we analyzed these corrected $EW$s; in the future papers of
this series, devoted to other clusters, we plan to check and calibrate 
our $EW$s by comparison with stars with both UVES and GIRAFFE spectra in a
systematic way.

Line lists and atomic parameters for the lines falling in the spectral range
covered by gratings HR11 and HR13 are from Gratton et al. (2003b) and are
described in Paper I, together with the adopted solar reference
abundances. For the interested reader, a more comprehensive discussion
of the atomic parameters and sources of oscillator strengths is provided
in Section 5 and Table 8 of Gratton et al. (2003b).

\section{Atmospheric parameters and analysis}

\subsection{Atmospheric parameters}

Temperatures and gravities were derived as described in Paper I; along with the derived atmospheric parameters and iron
abundances, they are shown in Table \ref{t:atmpar} (completely available only
in electronic form). We used $J,K$ magnitudes taken from the Point Source
Catalogue of 2MASS (Skrutskie et al. 2006); the 2MASS photometry was transformed to
the TCS photometric system, as used in Alonso et al. (1999).

We obtained $T_{\rm eff}$'s and bolometric corrections B.C. for our stars from
$V-K$ colors whenever possible.  We employed the relations by Alonso et al.
(1999, with the erratum of 2001). We adopted for NGC~6752 a
distance modulus of $(m-M)_V$=13.24, a reddening of $E(B-V)$ = 0.04,  an input
metallicity  of [Fe/H]$=-1.42$\footnote{This  value is somewhat different from
what we derive in the present study (i.e. $-1.56$), but the dependence of
($V-K$) on [Fe/H] is so weak that  temperatures are unaffected.} 
(from Gratton et al. 2003a), and the relations  $E(V-K) = 2.75 E(B-V)$, $A_V =
3.1 E(B-V)$, and $A_K = 0.353 E(B-V)$ (Cardelli et al. 1989). 

The final adopted  $T_{\rm eff}$'s were derived from a relation 
between $T_{\rm eff}$ (from $V-K$ and the Alonso et al.
calibration)
and $V$ magnitude based on 135 "well behaved" stars (i.e., with magnitudes in all the
four filters and lying on the RGB). The assumption behind this procedure
is that there is an unique relation between $V$ and $V-K$; this is a
sound assumption insofar (i) all stars belong to the RGB; (ii) the RGB is
intrinsically extremely thin (i.e. there is no spread in abundances) and (iii)
the reddening is the same for all the stars. There is no reason to question the
last point for NGC~6752 (the reddening itself is very small, see e.g. Gratton
et al. 2003a). The second point can only be verified a posteriori, by looking
at the derived abundances. Regarding the first hypothesis, the contamination of
stars on the AGB is not of concern in our sample, as discussed in the previous
section.
This procedure was adopted in order to
decrease the scatter in abundances due to uncertainties in temperatures, since
magnitudes are much more reliably measured than colours, hence the adopted
procedure produces extremely small nominal errors in T$_{\rm eff}$'s, see Sect. 4
below.

Surface gravities log $g$'s were obtained from effective temperatures and 
bolometric corrections, assuming  masses of 0.85
M$_\odot$ and  $M_{\rm bol,\odot} = 4.75$ as bolometric magnitude for the Sun.

\begin{figure}
\centering
\includegraphics[bb=30 150 510 700, clip, scale=0.55]{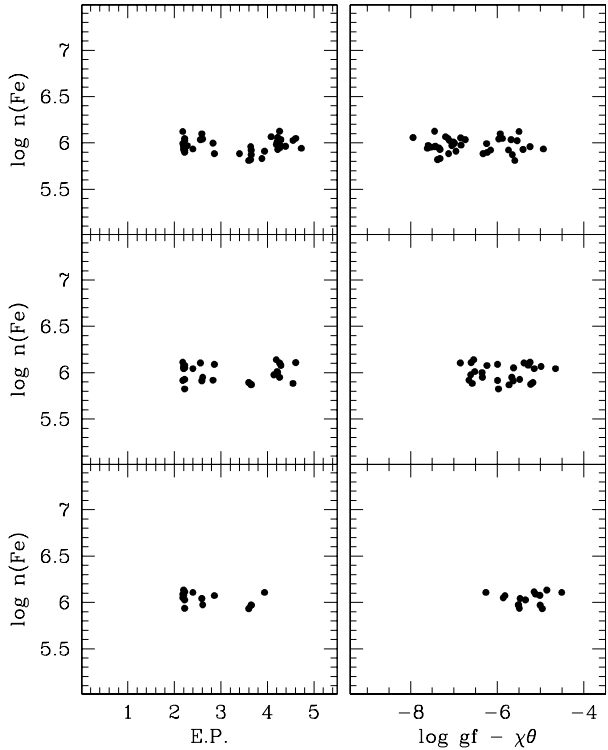}
\caption{Abundances deduced from neutral Fe {\sc i} lines as a function of
excitation potential (left panels) and of the expected line strength (right
panels) for three stars on the RGB of NGC~6752: a bright giant (star 48975,
T$_{\rm eff}=4337$~K, upper panels), a star in the middle of our temperature
range (star 1217, T$_{\rm eff}=4746$~K, middle panels) and a faint giant (star
24936, T$_{\rm eff}=4987$~K, lower panels).}
\label{f:figdew}
\end{figure}

We obtained values of the microturbulent velocity $v_t$ by eliminating
trends of the abundances from Fe {\sc i} lines with $expected$ line strength 
(see Magain 1984). The optimization was done for individual stars; this results
in a much smaller scatter in the derived abundances than using a mean relation
of $v_t$ as a function of temperature or gravity. Concerning stars observed
only with the grating HR11, since only a few lines where available,
we ended the optimization when the slope of
abundances from Fe {\sc i} lines vs expected line strength was within 1$\sigma$
error. Examples of the abundances from
neutral Fe lines as a function of the expected line strength and of the
excitation potential are shown in Figure~\ref{f:figdew} for three stars 
that cover the entire temperature range of our sample in NGC~6752.
Adopted atmospheric parameters are listed in Table~\ref{t:atmpar}.

Final metallicities are obtained by choosing by interpolation in the
Kurucz (1993) grid of model atmospheres (with the option for overshooting on)
the model with the proper atmospheric parameters whose abundance matches that
derived from Fe {\sc i} lines.

Average abundances of iron for NGC 6752 are [Fe/H]{\sc i}=$-1.56$ (rms=0.04 
dex, 137 stars) and [Fe/H]{\sc ii}=$-1.48$ (rms=0.09 dex, 105
objects). We do not think this difference is really relevant, since
abundances for Fe {\sc ii} rely on average on only two lines. 
Derived Fe abundances are listed in Table~\ref{t:atmpar}. 

\begin{figure}
\centering
\includegraphics[bb=110 210 540 710, clip, scale=0.65]{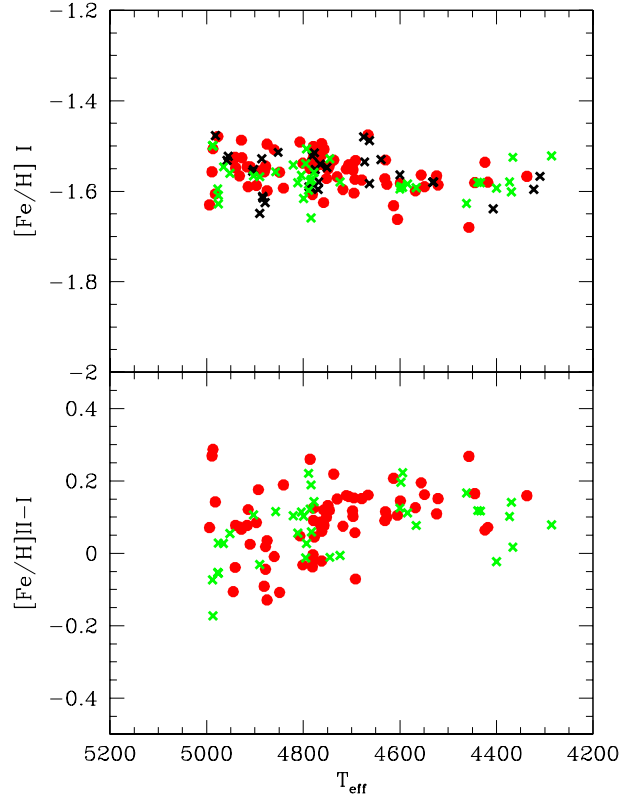}
\caption{Run of [Fe/H] ratio and of the iron ionization equilibrium as a
function of temperatures for program stars in NGC 6752. Symbols and color
coding refer to the setup used: (red) filled circles indicate stars with 
both HR11 and HR13 observations,
(black) crosses for HR11 only, and (green) crosses for HR13 only.}
\label{f:feteff}
\end{figure}

The distribution of the resulting [Fe/H] values and of the difference between
ionized and neutral iron are shown in Figure~\ref{f:feteff}, as a function of
temperature, with stars coded according to the grating(s) they were observed
with. The (small) scatter of the metallicity distribution is discussed 
in Sect. 5.

Total errors, computed using only the dominant terms or including all the
contributions (see Sect. 4), are reported in 
Table~\ref{t:sensitivity}, in Cols. 8 and 9 respectively. They were
scaled down by weighting the sensitivity abundance/parameter
with the actual internal error in each parameter.

As already mentioned in the Introduction, the metallicity of NGC~6752 has  been
determined by several other authors. Limiting to very recent papers that
analyzed UVES spectra of resolution higher (R=45000) or even much higher
(R=60000 or 110000) than ours, we cite: 
Gratton et al. (2001), based on spectra of main sequence and subgiant branch stars
([Fe/H]=$-1.42$);
Gratton et al. (2005), based on FLAMES/UVES spectra of 7 RGB stars near to the
RGB bump ([Fe/H]{\sc i}=$-1.48$, [Fe/H]{\sc ii}= $-1.55$);
Yong et al. (2005), based on very high resolution spectra of 38 RGB stars
([Fe/H]=$-1.61$).
A detailed comparison with literature results would require to understand all
the systematics between different analyses (e.g., temperature scale, model
atmospheres, etc) and 
is completely outside the scope of the present paper.
Overall, if we limit ourselves to red giants
the agreement is excellent, showing the reliability of iron abundances
derived from relatively low resolution spectra as those from the GIRAFFE/MEDUSA
instrument, once they are checked and calibrated on the scale of $EW$s
derived from the higher resolution UVES spectra using stars observed in both
modes. The difference is larger when abundances derived from dwarfs and early
subgiants are considered. While part of this difference can be related to the
different line subset used in the analyses, it is possible that the discrepancy
is caused by systematic differences in the atmospheres of dwarfs and giants
with respect to the models by Kurucz.

\begin{table*}
\centering
\caption[]{Adopted atmospheric parameters and derived iron abundances in 
stars of NGC 6752; nr indicates the number of lines 
used in the analysis. 
The complete Table is available only in electronic form.
}
\begin{tabular}{rccccccrccc}
\hline
Star   &  $T_{\rm eff}$ & $\log$ $g$ & [A/H]  &$v_t$         & nr & [Fe/H]{\sc i} & $rms$ & nr &[Fe/H]{\sc ii} & $rms$ \\
       &     (K)        &  (dex)     & (dex)   &(km s$^{-1}$) &    & (dex)         &       &    & (dex)         &       \\
\hline
   1217  & 4746 &2.14& $-$1.55 &1.50 & 26&   -1.54& 0.09&     3 & -1.42& 0.09	\\  
   1493  & 4821 &2.30& $-$1.55 &1.33 & 16&   -1.54& 0.07&     2 & -1.44& 0.04	\\  
   1584  & 4916 &2.48& $-$1.55 &1.41 & 14&   -1.55& 0.08&     1 & -1.47&	\\  
   1797  & 4804 &2.25& $-$1.58 &1.49 & 20&   -1.57& 0.08&     2 & -1.45& 0.01	\\  
   2097  & 4418 &1.41& $-$1.59 &1.71 & 39&   -1.58& 0.09&     3 & -1.51& 0.07	\\  
   2162  & 4693 &2.02& $-$1.58 &1.52 & 26&   -1.57& 0.06&     2 & -1.52& 0.06	\\  
   2704  & 4786 &2.20& $-$1.56 &1.47 & 20&   -1.55& 0.07&     3 & -1.29& 0.08	\\  
   2817  & 4841 &2.31& $-$1.61 &1.60 & 22&   -1.59& 0.11&     2 & -1.40& 0.03	\\  
   4602  & 4801 &2.24& $-$1.52 &1.33 & 20&   -1.54& 0.11&     1 & -1.57&	\\  
\hline
\end{tabular}
\label{t:atmpar}
\end{table*}

\begin{table*}
\centering
\caption[]{Sensitivities of abundance ratios to variations in the atmospheric
parameters and to errors in the equivalent widths, as computed for a typical 
program star with $T_{\rm eff} \sim 4750$ K. 
The total error is computed as the quadratic sum
of the three dominant sources of error,
$T_{\rm eff}$, $v_t$ and errors in the $EW$s, scaled to the actual errors as
described in the text (Col. 8: tot.1) or as the sum
of all contributions (Col. 9: tot.2)}
\begin{tabular}{lrrrrrrrr}
\hline
\\
Ratio    & $\Delta T_{eff}$ & $\Delta$ $\log g$ & $\Delta$ [A/H] & $\Delta v_t$
&$<N_{lines}>$& $\Delta$ EW & tot.1 & tot.2\\
         & (+50 K)    & (+0.2 dex)      & (+0.10 dex)      & (+0.10 km/s) & & & (dex)& (dex)  \\
 (1) & (2) & (3) & (4) & (5) & (6) & (7) & (8) & (9) \\
\\
\hline
$[$Fe/H$]${\sc  i} &  +0.063 & $-$0.007 &$-$0.009 & $-$0.025 & 23 &+0.020  &0.039 &0.039  \\
$[$Fe/H$]${\sc ii} &$-$0.024 &   +0.085 &  +0.023 & $-$0.010 &  2 &+0.068  &0.069 &0.070  \\
\hline
$[$O/Fe$]$         &$-$0.046 &   +0.088 &  +0.026 &   +0.023 &  1 &+0.096  &0.101 &0.102  \\
$[$Na/Fe$]$        &$-$0.023 & $-$0.029 &$-$0.004 &   +0.018 &  3 &+0.055  &0.060 &0.060  \\
\hline
\end{tabular}
\label{t:sensitivity}
\end{table*}

\section{Errors in the atmospheric parameters}

In this section we will consider possible analysis errors leading to an
increase in the scatter of the relations used throughout this paper. We leave
aside errors arising from the simplification in the analysis (1-d
plane-parallel model atmospheres, LTE approximation etc.). These errors are
relevant when comparing abundances of stars in different evolutionary phases
and with different atmospheric parameters (see e.g., Asplund 2005); this might possibly
be an explanation of the differences between abundances for dwarfs and red
giants, which appears to be rather large and significant. However, we expect
such errors to show up as trends in the derived abundances, rather than in
scatter at a given location in the color-magnitude diagram. Possibly, such
errors might be responsible, e.g., for the systematic trend for the coolest
stars to produce lower values of the Fe abundances from neutral lines. However,
this effect is small within the range of temperatures (from 5000 to 4300 K)
considered here. 

Errors in the derived abundances are mainly due to three main sources, i.e.
errors in temperatures, in microturbulent velocities and in the
measurements of $EW$s. Less severe are the effects of errors in
surface gravities and in the adopted model metallicity.
As in Paper I, in the following we will concentrate on the 
major error sources.

\paragraph{Errors in temperatures.} 
The final derivation of temperatures in NGC~6752 slightly differs from the
procedure adopted for NGC~2808 (Paper I). There is no evidence of differential
reddening in NGC~6752, hence for the stars in the present sample we adopted a
final $T_{\rm eff}$ value from a calibration of temperature as a function of
the $V$ magnitude. This approach might not be the best for the coolest stars
(namely those with $T_{\rm eff} < 4500$~K), where some degree of variability
in magnitude may be expected, due to the small number of ascending
convective cells in the atmosphere. However, this effect should be amply
compensated by the increase of the $S/N$ ratio.

The nominal internal error in $T_{\rm eff}$ is estimated
 from the
adopted error of 0.02 mag in $V$ (a conservative estimate, very likely 
overestimated, in this magnitude range, where photometric errors are less than
a few thousandths of magnitude, see Momany et al. 2004) and the slope of the 
relation between temperature and magnitude. In NGC~6752 this slope is on
average 248 K/mag, hence the conservative estimate of the star-to-star errors
in the $V$ magnitude of 0.02 mag produces an internal error as low as 5~K.
Of course, systematic and in particular scale errors might be considerably
larger. We want to stress here that this
is the error relevant when we intend to study a star-to-star variation in some
elemental abundances in a single cluster, as is the present case for the Na-O
anticorrelation (see below for a discussion of the meaning we
attribute to these small errors).

\paragraph{Errors in microturbulent velocities.}
We used star 1217 and repeated the analysis changing $v_t$ until the 1$\sigma$
value\footnote{
This value was derived as the quadratic mean of the 1 $\sigma$ errors in the
slope of the relation between abundance and expected line strength for all stars
with more than 15 lines measured.} 
from the original 
slope of the relation between line strengths and
abundances was reached; the corresponding internal error is 0.13 km~s$^{-1}$. 
This quantity is larger than the one we found for NGC~2808 (Paper I: 0.09
km~s$^{-1}$), since it depends on the position of the measured lines along the
curve-of-growth: in NGC 6752 we measured a smaller number of lines, 
which are moreover weaker, hence less sensitive to the microturbulence.
The scatter in individual microturbulent velocities along a mean regression
line with gravity (or effective temperature) is of 0.17 km~s$^{-1}$ (stars with
more than 15 lines measured) or 0.21 km~s$^{-1}$ (all stars). Scatter in Fe
abundances are reduced when adopting individual values, with respect to the
result obtained using the mean line. This strongly suggests that this scatter
reflects a real characteristic of the atmospheres, rather than simply an
observational error. On the other hand, the physical interpretation of the
``microturbulent velocity" in an analysis like ours is not straightforward.
While it might indicate the velocity fields in the stellar atmosphere, it might
also be related to a different (average) dependence of the temperature on the
optical depth, since microturbulent velocities are derived by eliminating trends
in abundances with line strength rather than by fitting line profiles (a
procedure that would however be meaningless at the resolution of our spectra).
Hence, the most correct interpretation of the observed scatter in the
microturbulent velocities is that the atmosphere of each star has its own
differences with respect to the average model atmosphere, either in the velocity
field, or in the thermic structure, or both. It is also possible that these
differences change with time, and that a spectrum obtained at a different
epoch for the same star would be interpreted with a different microturbulent
velocity. This is in fact what should be expected if convection is a
time-variable phenomenon, as expected by 3-d model atmospheres (see Asplund
2005) taking into account the large size of the convective cells in the
atmospheres of red giants.  On this respect, the very small nominal error we
obtained for the effective temperatures should not be overinterpreted.

\paragraph{Errors in measurement of equivalent widths.}
Errors in the individual $EW$s may be estimated by comparing the $EW$s
measured in couples of stars with very similar atmospheric parameters and S/N
ratios,  and computing the rms scatter about the linear relationship between
the two stars. Assuming that both sets of $EW$s have equal errors, we may
estimate the typical internal errors in $EW$s. We applied this procedure to
pairs of stars over all the temperature range of our sample. Typical average
measurement errors of 3.7~m\AA\ were derived, yielding typical line-to-line
scatter in abundances of 0.096 dex for Fe {\sc i} (this is the quadratic mean
over 92 stars with more than 15 measured Fe lines). The contribution of random
errors in the $EW$s to errors in the abundances can then be obtained by
dividing this typical errors for individual lines by the square root of the
typical number of lines used in the analysis (22), providing a typical internal 
error of 0.020 dex.

\paragraph{}
Once we have derived the internal errors we may compute the final errors in
abundances; they depend on the slopes of the relations between the variation
in each given parameter and the abundance. 
Columns from 2 to 5 of Table~\ref{t:sensitivity} show the sensitivity of the 
derived abundances to
variations in the adopted atmospheric parameters for Fe, Na and O. This 
has been
obtained by re-iterating the analysis while varying each time only one of the
parameters of the amount shown in the Table.
This exercise was done for all stars in the sample. The average
value of the slope corresponding to the average temperature ($\sim 4750$ K) in
the sample was used as representative to estimate the internal errors 
in abundances, according
to the actual uncertainties estimated in the atmospheric parameters.  For
iron, these amount to $\sim 0.01$ dex and 0.033 dex, due to the quoted 
uncertainties of 5~K and 0.13 km/sec in $T_{\rm eff}$ and $v_t$.
The impact of errors in $EW$s is evaluated in  Col. 7, where the average error
from a single line is weighted by the square root of the mean number of lines,
given in Col. 6. This has been done for iron and for the other two elements
measured in this paper.
Of course this approach corresponds to an interpretation of the variation
of microturbulence as due to real velocity fields in the stellar atmosphere,
and to the adoption of the nominal values for the errors in the effective
temperatures. While the above discussion shows that this interpretation may be
incorrect, we notice that the good agreement with the actual scatter in Fe
abundances (see next section) suggests that the total (internal) error given 
by this interpretation is not grossly wrong.

\section{Results}
\subsection{Cosmic scatter and reddening}

To evaluate the expected scatter in [Fe/H] due to the uncertainties in
$T_{\rm eff}$, $v_t$ and errors in $EW$s we make
use of Table~\ref{t:sensitivity}; we
derive $\sigma_{\rm FeI}$(exp.)=$0.039 \pm 0.003$ dex (statistical error). The
inclusion of contributions due to uncertainties in surface gravity or model
metallicity does not alter our conclusions. 
The observed scatter, 
estimated as the average rms scatter that we obtain
using the 92 stars in our sample with at least 15 measured iron lines, after a
3$\sigma$ clipping, is formally lower:
$\sigma_{\rm FeI}$(obs.)=$0.037 \pm 0.003$ (statistical error).
However, within the statistical uncertainties this difference is not
significant, and might indicate that the errors are slightly overestimated.
The conclusion from our large dataset is that 
the observed star-to-star rms scatter in Fe abundances in NGC 6752 is no
more than 9\% (decreasing to 8\% had we adopted a more tight clipping at
2.5$\sigma$).

This is the second cluster of our sample and we can check 
that we are obtaining results as homogeneous as possible.
One can worry e.g., about the fact that we are deriving \teff's and $\log g$'s 
from the photometry using reddening values and distances coming from
different sources for individual clusters.

Distance has a direct impact on $\log g$, but reasonable systematic
errors on it (of the order of 0.2 mag in distance modulus) translate
into a difference of less than 0.1 in $\log g$, which is negligible, at least
for Fe {\sc i} (see Table \ref{t:sensitivity}). An even more negligible
effect is produced by systematic differences in the reddening scales
between the two analyses.   

On the contrary, the influence of diverse reddening scales on the temperatures
is noticeable: a systematic shift of 0.02 mag means \teff's from $V-K$
differing by about 40~K, or Fe {\sc i} abundances differing by about 0.05 dex.
We may estimate the consistency of the reddening values adopted in  the two
cases (0.22 and 0.04 for NGC~2808 and NGC~6752, respectively) comparing two
different temperatures scales, one dependent from the reddening (\teff's
derived from the photometry) and one independent from it (\teff's derived from
the line excitation equilibrium).
Let $\Delta \theta = \Delta({\rm Fe I}/ EP)$ be the slope of the relation
between abundances from neutral iron lines and excitation potential; we can
derive a relation between $\Delta \theta$ and the photometric temperature, and
we obtain that for temperatures (based on $V-K$) of about 4500~K, a difference of
45~K corresponds to $\Delta \theta = 0.013$ dex/eV.
The model atmospheres and the oscillator strengths used in the two analyses
are the same, so they do not introduce differences; if the reddenings are
on the same scale, the average $\Delta \theta$'s
for the two clusters should have to be the same.
We chose only stars with enough measured lines to derive \teff \
spectroscopically, i.e. those with $>25$ lines in NGC~2808 (there are 63)
and $>15$ lines in NGC~6752 (there are 92), obtaining
$\langle \Delta \theta \rangle(2808) = -0.009 \pm 0.003$ (rms=0.026) and
$\langle \Delta \theta \rangle(6752) = -0.006 \pm 0.003$  (rms=0.026).
Given the above dependence of $\Delta \theta$ on \teff, the difference
in $\Delta \theta$'s between the two clusters ($-0.003 \pm 0.004$)
corresponds to -10~K, or -0.004 mag in \ebv.
Furthermore, 10~K correspond to slightly more than 0.01 dex in Fe {\sc i}.

We may safely assume that the two reddening values and metallicities are
on a consistent scale.

\subsection{The Na-O anticorrelation}

Abundances of O and Na rest on measured $EW$s (or upper limits). 
Abundances of Na could be measured for all stars; depending on the setup 
used, at least one of the Na {\sc i} doublets
at 5672-88~\AA\ and at 6154-60~\AA\  is always 
available. Derived average Na abundances were corrected for effects of
departures from the LTE assumption using the prescriptions by Gratton et al.
(1999).

Oxygen abundances are obtained from the forbidden [O {\sc i}] lines at 6300.3
and 6363.8~\AA; the former was cleaned from telluric contamination by
H$_2$O and O$_2$ lines before extracting the EW.

CO formation is not a source of concern in the derivation of O abundances
due to the low expected C abundances
and to the rather high temperatures of these stars. Also, as in Paper I, we
checked that the high excitation Ni {\sc i} line at 6300.34~\AA\ is not a
substantial contaminant, giving a negligible contribution to the measured
$EW$s of the forbidden [O {\sc i}] line.

Abundances of O and Na are listed in Table~\ref{t:abunao} (the complete table
is available only in electronic form). For O we distinguish between actual
detections and upper limits; for Na we also
indicate the number of measured lines and the rms value.

\begin{figure}
\centering
\includegraphics[bb=25 170 410 700, clip, scale=0.70]{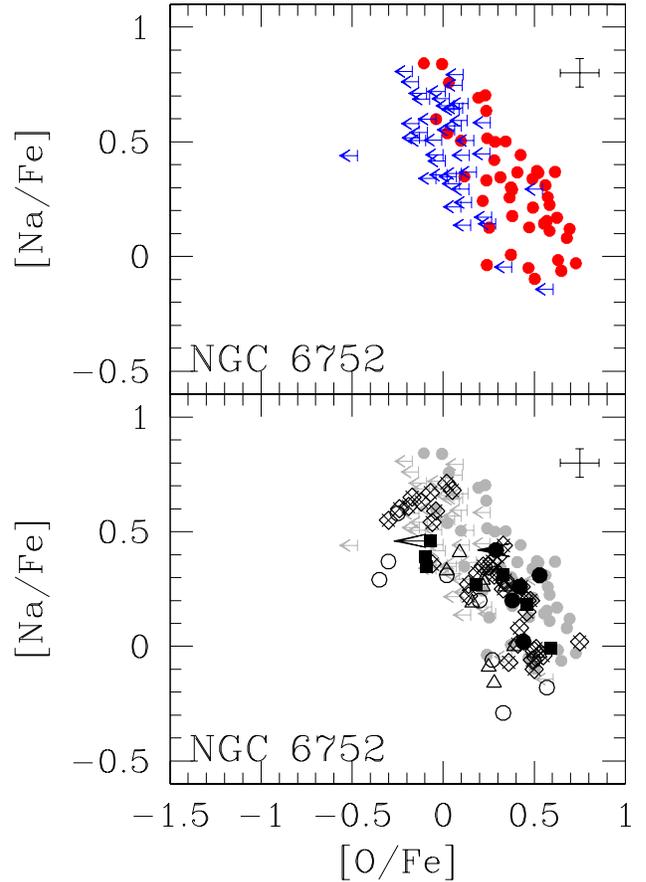}
\caption{Upper panel: [Na/Fe] ratio as a function of [O/Fe] for red giant 
stars in NGC 6752 from the present study. Upper limits in [O/Fe] are indicated
as blue arrows. The error bars take into account the uncertainties in
atmospheric parameters and $EW$s. Lower panel: literature data from several
study (see text) superimposed to our results. Filled and open large circles
are subgiant and turnoff stars from Gratton et al. (2001) and Carretta et al.
(2004). Filled squares are RGB stars from Gratton et al. (2005). Diamonds with
crosses inside are RGB stars from Yong et al. (2003, 2005). Open triangles are
giants from Norris and Da Costa (1995) and Carretta (1994).}
\label{f:anti}
\end{figure}

The [Na/Fe] ratio as a function of [O/Fe] ratio is displayed (filled dots) in
the upper panel of Figure~\ref{f:anti} for each of the red giant stars with 
both O and Na detections in NGC 6752; stars in which only upper limits in the
$EW$s of the [O {\sc i}] 6300~\AA\ line were measured are also shown as
arrows.  Despite
the moderately high  resolution and  relatively high S/N ratios of our spectra we were not
able to reach down to [O/Fe]$\sim -1$, at variance with NGC~2808 (Paper I)
where we measured such low abundances, thanks to the high quality of
the spectra and to the higher cluster metallicity. 

\begin{figure}
\centering
\includegraphics[bb=55 196 420 500, clip, scale=0.70]{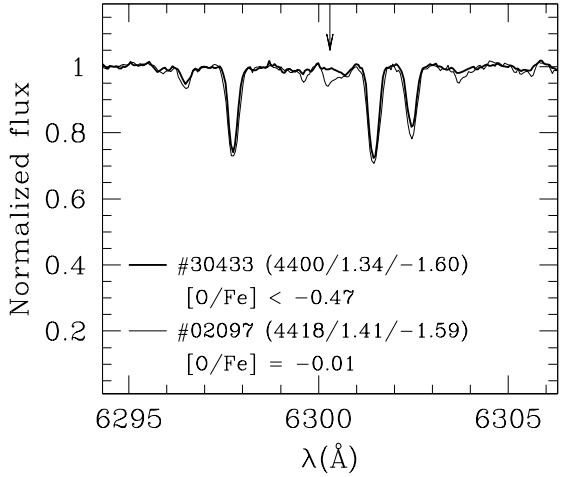}
\caption{Comparison of the observed spectra of stars 30433 and 2097 in 
NGC 6752 near
the [O {\sc i}] 6300.31~\AA\ line. These stars have very similar
atmospheric parameters ($T_{\rm eff}$, $\log g$ and [Fe/H] are indicated), 
yet quite different [O/Fe] abundances.}
\label{f:confspec}
\end{figure}

However, from our study alone we cannot completely exclude that a few very
O-poor stars are  also present in this cluster.
A possible example is provided by star 30433, for which we only derived an 
upper limit [O/Fe]$< -0.53$ dex. In Figure~\ref{f:confspec} we show the
spectral region including the forbidden [O {\sc i}] 6300.31~\AA\ line for 
this star, compared to
the same region  star 2097, with very similar atmospheric parameters.
For star 2097 (with an estimated $S/N\sim 250$) we actually measured an $EW$
whereas only a conservative upper limit can be safely derived for star 30433.
Notice that the estimated $S/N$ ratio for this latter is about 135 per
pixel, hence we are quite confidant that the lack of measurable features at 
6300~\AA\ is real.

However, the overall impression is that stars with very low O abundances
([O/Fe]$<-0.5$ dex) are quite scarce in this cluster, if
any.
This conclusion is supported by the comparison with other literature 
results for NGC 6752, shown in the lower panel of Figure~\ref{f:anti}.
In this panel we superimposed to the present data O and Na abundances from a
number of studies, derived both from giants and unevolved stars (from
Gratton et al. 2001 and Carretta et al. 2004 for subgiant and dwarf stars; from
Gratton et al. 2005, Yong et al. 2003, 2005, Norris \& Da Costa 1995, 
and Carretta 1994 for RGB stars). 
The agreement with previous analyses is very good.
All these studies are based on high resolution spectra, and
in none of these an extremely O-poor star is present, not even in the
dataset by Yong et al. (2003, 2005), where very high S/N values (up to 300-400) 
are obtained.

NGC 6752 seems in this respect more similar to the majority of GCs 
(see Fig. 5 in Paper I)
where the bulk of measured [O/Fe] ratios is in the range +0.5 to $-0.5$.

\begin{table}
\centering
\caption[]{Abundances of O and Na in NGC 6752. [Na/Fe] values are corrected for
departures from LTE. HR is a flag for the grating used 
(1=HR13 only, 2=HR11 and HR13, 3=HR11 only) and lim is a flag discriminating
between real detections and
upper limits in the O measurements (0=upper limit, 1=detection). The complete
Table is available only in electronic form.}
\begin{tabular}{rrccrcccc}
\hline
Star     &   nr &  [O/Fe] & rms   &   nr &   [Na/Fe]&  rms   & HR & lim \\
\hline
   1217   & 1&   +0.01&     &  4 &  +0.72& 0.07 & 2 & 0 \\  
   1493   & 1&   +0.02&     &  2 &  +0.36& 0.09 & 1 & 0 \\  
   1584   & 1&   +0.29&     &  2 &  +0.14& 0.08 & 2 & 0 \\  
   1797   & 1&   +0.01&     &  2 &  +0.42& 0.09 & 1 & 0 \\  
   2097   & 2& $-$0.01& 0.03&  3 &  +0.84& 0.05 & 2 & 1 \\  
   2162   & 1&   +0.10&     &  4 &  +0.57& 0.06 & 2 & 0 \\  
   2704   & 1&   +0.58&     &  3 &  +0.26& 0.05 & 2 & 1 \\  
   2817   & 1&   +0.62&     &  3 &  +0.37& 0.14 & 2 & 1 \\  
   4602   & 1& $-$0.01&     &  3 &  +0.51& 0.11 & 2 & 0 \\  
   4625   & 1&   +0.47&     &  2 &$-$0.05& 0.10 & 2 & 1 \\  
   4787   & 1&   +0.11&     &  3 &  +0.65& 0.05 & 2 & 0 \\  
\hline
\end{tabular}
\label{t:abunao}
\end{table}

\begin{figure}
\centering
\includegraphics[bb=20 160 320 700, clip, scale=0.80]{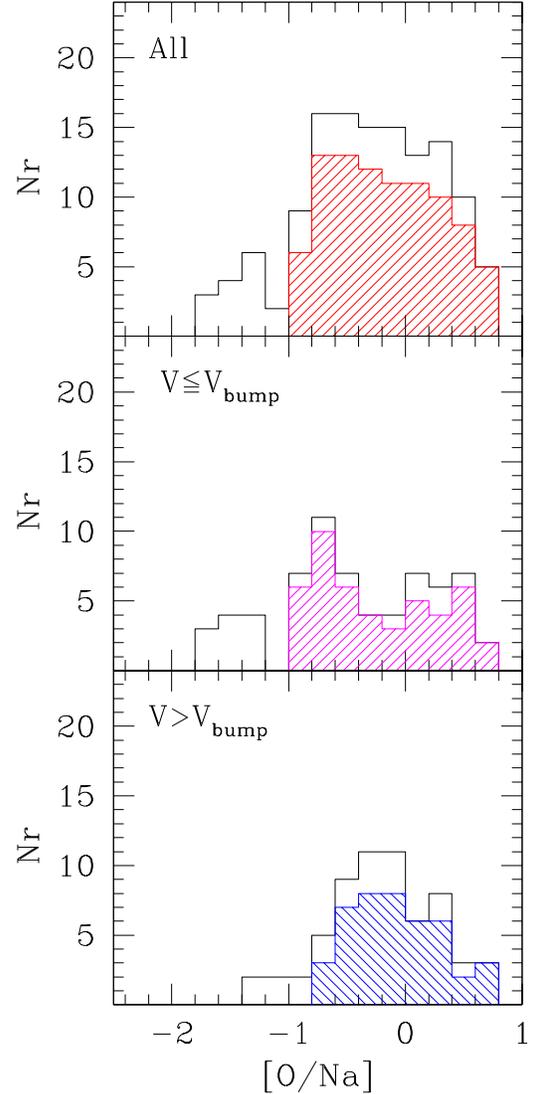}
\caption{Upper panel: distribution function of the [O/Na] ratios along 
the Na-O anticorrelation in NGC 6752. The dashed area is the frequency histogram
referred to actual detection of O in stars, whereas the empty histogram is
obtained by using the global anticorrelation relationship derived in Paper I
 to obtain O abundances
also for stars with no observation with HR13. 
Middle and lower panel: the same, for stars brighter and fainter
than the magnitude of the RGB-bump in NGC 6752 ($V=13.65$), respectively.}
\label{f:histom67}
\end{figure}

\begin{figure}
\centering
\includegraphics[bb=55 200 390 700, clip, scale=0.70]{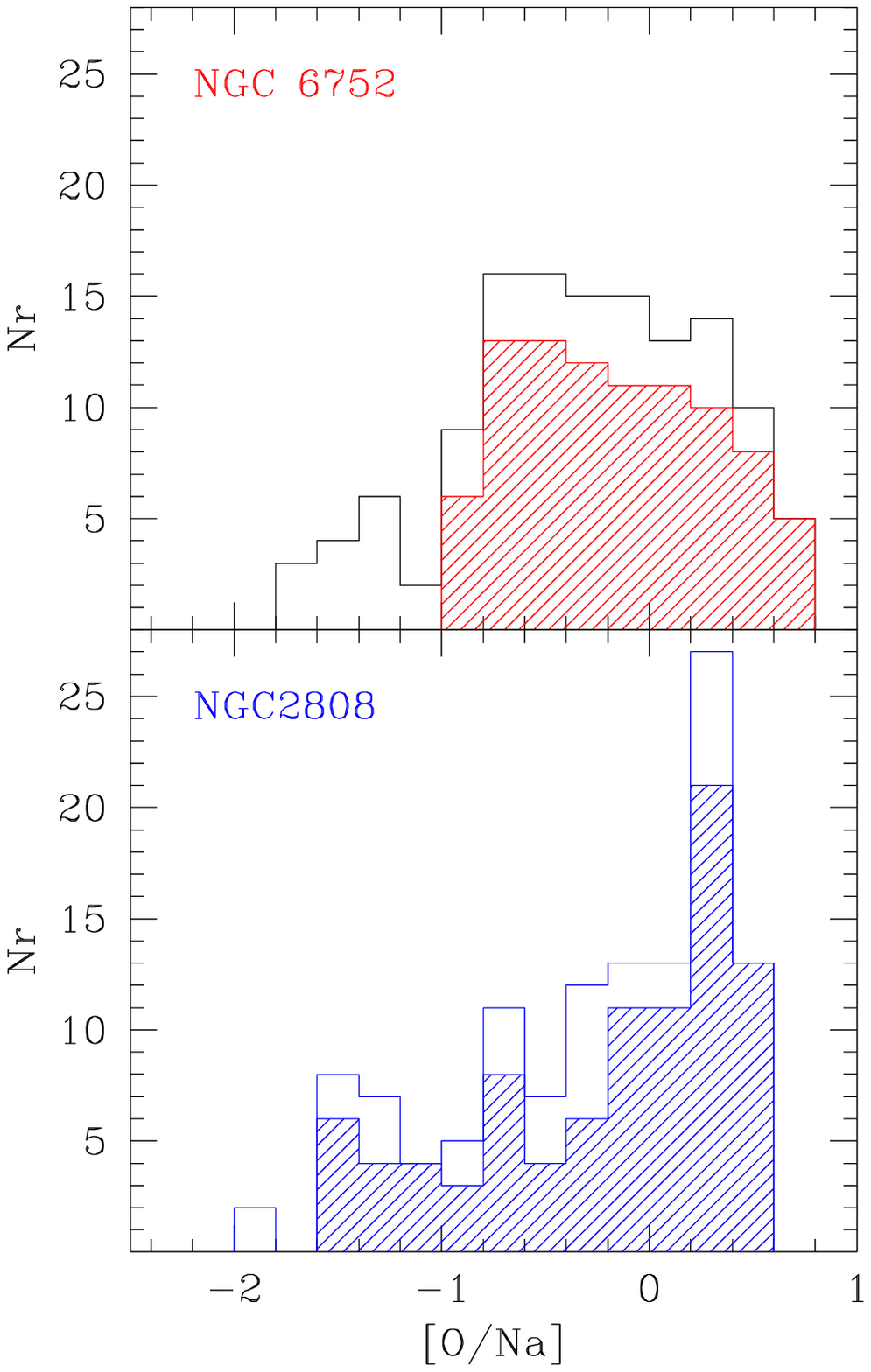}
\caption{Comparison of the distribution functions of the [O/Na] ratios 
for stars along the red giant branches in NGC 6752 (upper panel, this work) and
in NGC 2808 (lower panel, Carretta et al. 2006).}
\label{f:histom67m28}
\end{figure}

The distribution function of stars along the Na-O anticorrelation in NGC 6752
is shown in the upper panel of Figure~\ref{f:histom67}, where the ratio [O/Na] 
from our data is used. The dashed area shows the distribution obtained by using only actual
detections or carefully checked upper limits. The empty histogram is derived by
following the overall Na-O anticorrelation, as derived in Paper I  from a
collection of literature data and NGC~2808, in order to estimate [O/Fe] values
even for stars with no observations in HR13.

The middle and lower panels of Figure~\ref{f:histom67} show the distribution
functions for stars brighter and fainter than the magnitude level of the
RGB-bump ($V=13.65$, estimated from the photometry by Momany et al. 2004; the
bump is also clearly visible in Figure~\ref{f:figcmd}). 
From a Kolmogorov-Smirnov statistical test the two distributions 
may be extracted from the same parent distribution. 
The magnitude of the bump on the RGB marks the evolutionary point where
a second phenomenon of mixing ("extra-mixing") is allowed to onset in
Population II red giants (see Gratton et al. 2000 for a detailed discussion and
references). Our results, based on the largest sample ever studied with
homogeneous procedures in this
cluster, strongly support the conclusion that in NGC 6752 the
bulk of chemical anomalies in Na and O abundances is primordial, already 
established in stars before any evolutionary mixing is allowed. 
Any further modification in the [Na/Fe] or [O/Fe] ratios due to mixing
mechanisms during the evolution along the RGB must be considered as a second 
order effect.

\section{Discussion}

The first conclusion of our study is that in NGC~6752 there is no measurable intrinsic spread in
metallicity, and that this cluster is  very homogeneous as far as the global
metallicity is concerned. In turn, this result emphasizes that every model
proposed to explain the formation of a globular cluster has to face very tight
constraints. Abundance anomalies and anticorrelations
between elements forged in high temperature $p$-capture reactions are likely
the relics of ejecta from intermediate-mass 
AGB stars\footnote{
Very recently, Prantzos and Charbonnel (2006) suggested instead 
that the Na-O anticorrelation is due to nucleosynthesis in rapidly rotating
massive stars rather than in AGB stars. While this difference would not affect
most of the present discussion, we note that the very small scatter in Fe
abundances sets stringent limits on the ``inefficiency" of mixing between the
slow winds of stars in the supergiant phases and the ejecta from core-collapse
supernovae in this scenario.}. 
On the other hand, iron
is not produced in such stars, but mainly in more massive stars exploding early
as core-collapse supernovae. 
We might expect that the proto-GCs might have been able to have some
independent chemical evolution, retaining at least part of the metal-enriched
ejecta of core-collapse SNe (Cayrel 1986; Brown et al. 1991, 1995; 
Parmentier \& Gilmore 2001).
Considering the typical mass of a globular cluster
(likely larger in the past, due to possible stripping and evaporation of stars)
some tens of supernovae are presumably required in order to enrich the whole
protocluster. 
Thus, a physically realistic model $must$ include
a very efficient (turbulent?) mixing of the protocluster cloud, in order to
match the very homogeneous metallicity of the observed stars in NGC 6752.

The second item to be considered concerns the Na-O anticorrelation.
We have currently available 2 GCs with much more than 100 stars observed and
analyzed in a very homogeneous way (the largest samples ever
used to study in detail the Na-O anticorrelation). Therefore, it is
tempting to draw some preliminary inferences from our datasets.
When coupled with other literature data, some simple statements are already
possible. In fact, we know that:
\begin{itemize}
\item[(i)] as Sneden et al. (2004) pointed out, red giants in NGC 6752 and M 13
share the same anticorrelation between [O/Fe] and the same relative
ratio of the Mg
isotopes. On the other hand, the excellent study by Yong et al. (2003) in NGC
6752 shows that the source of the Mg isotopes is likely a generation of
intermediate mass stars in the range 3-6 M$_\odot$, confirming the earlier
findings by Gratton et al. (2001) from the abundance analysis of unevolved
stars in this cluster.
\item[(ii)] the comparison between NGC 6752 and M 13 seems to imply that the
latter suffered a greater degree of pollution than the former.
\item[(iii)] the HB morphology in NGC 6752 and M 13 is similar: the 
horizontal-branch ratio, HBR = (B-R)/(B+V+R), is 0.97 for
M~13 and 1.0 for NGC~6752, meaning that both M~13 and
NGC~6752 have blue HBs, with stars populating only the region hotter than
the RR Lyrae instability strip; 
\item [(iv)] large degree of anomalies as in O, Mg depletion and Na, Al
enhancement are observed also in NGC 2808 (Paper I and Carretta 2006), and they
are as extreme as those found in M 13; yet, the HBR parameter is  $-0.49$
for NGC~2808, which represents a more complicate case (Paper I and literature
there cited);
\item[(v)] however, Carretta (2006) convincingly showed that no correlation
seems to be present between the extension of both the Na-O and Mg-Al
anticorrelations and the HB morphology.
\end{itemize}

Figure~\ref{f:histom67m28} compares the results obtained from our large
datasets in NGC~6752 and NGC~2808.
The distribution functions look somewhat different; the Kolmogorov-Smirnov 
statistic returns a probability of only about 7\% that the two subsamples
are extracted from the same parent population. The distribution of stars along
the [O/Na] anticorrelation in NGC 6752 
is more similar to the one for M~13 than to the NGC~2808 one
(see Fig. 7 in Paper I). Although no super O-poor stars are found in the present
study or in any other published to date, there is 
still the possibility that NGC~6752 could host a handful of such stars.
In the moderately limited sample by Yong et al. (2003,2005), where
the O abundances are all detections based on high quality spectra, 
there are no super O-poor stars
out of 38 objects: according to the binomial distribution, 
at 95\% level of confidence a firm upper limit to the population of extremely
O-poor stars in NGC~6752 is 8\%. Regarding our much larger sample, this estimate
is more uncertain, due to the upper limits in the [O/Fe] ratios; however, the
results are not at odds with those by Yong et al. 

These findings emphasize a potential inconsistency in the scenario proposed
e.g. by D'Antona and Caloi (2004): if the progeny of super O-poor stars on the
RGB is located on the bluest part of the HB (EBHB), we should 
expect that less than about 8\% of the horizontal branch stars in NGC~6752
populate  this region of the HB. However, the populous blue tail of the HB in
this cluster contains at least about 20\% of HB stars, implying a discrepancy
of about a factor of 2. A possible way out is if most of EHB's are the outcome
of binary star evolution; however, this is at odds with results
by Moni Bidin et  al. (2006) who detected no close binary system among 51 hot
HB stars.
Moreover, in NGC~6752 also the progeny of ``normal", O-rich stars is confined
to the blue HB only. A detailed modeling and fine tuning of the mass
loss process would be required to reproduce the HB distribution in this
cluster.
 
Another possibility is that we are seeing less stars on the RGB than expected,
in particular those with extremely low O abundances, because they never
reach the upper part of the giant branch: these objects could maybe loose a
large amount of mass, becoming RGB-manqu\`e and ending up on the EBHB.
This hypothesis can be checked using
number counts of stars populating different parts of the CMD (RGB, HB) that
should be clearly affected. In particular, the R parameter, usually employed to
determine the helium abundance (e.g. Buzzoni et al. 1983) is based on the the
ratio of HB to RGB stars.
Interestingly enough, the latest
compilation of R parameters by Sandquist (2000) shows that 
several clusters with the bluest HB morphologies have unusually high R values,
NGC~6752 among these.
Unfortunately, nothing conclusive can be derived from the presently extant 
data: Sandquist reports a value R=1.56$\pm 0.18$ based on the rather old
photographic photometry by Buonanno et al. (1986). 
Thus the R parameter goes in the right direction, being able to explain the
excess of about 12\% of stars in the blue tail required to reconcile the
numbers of RGB-manqu\`e and HB stars, 
since it is about 12\% larger in this cluster, with respect to the
average of other GCs. However, the attached error is currently still too high 
(it is itself about 12\%) and precludes a firmer conclusion.

Finally, we note here that NGC~6752 and M~13 have similar metallicities, while NGC~2808
is slightly more metal-rich; the latter is also younger (Rosenberg et al. 1999)
than the bulk of GCs.  For NGC~6752 the HB populated only in the blue part must
be mostly due to the influence of the first parameter (metallicity) and maybe
to the age.

Further discussion is obviously postponed until a substantial fraction of our
sample of GCs, with different HB morphologies, has been studied, but the
emerging scenario is a rather complex one where several factors 
(excess of He from a prior
generation of stars, age, metallicity) may contribute to build up both the
final Na-O anticorrelation on the RGB and the star distribution on the HB. 
The relative weight of these contributions, as well as the global properties 
such as the cluster orbital parameters (see Carretta 2006), may then compete
to give the observed differences in individual objects.

\section{Summary}

In this paper we have derived atmospheric parameters and elemental abundances
for about 150 red giant stars in the globular cluster NGC 6752 observed with the
multifiber spectrograph FLAMES.

Atmospheric parameters for all targets were obtained from the photometry.
From the analysis of the GIRAFFE spectra we derived an average metallicity 
[Fe/H]$=-1.56$ (rms=0.04 dex, 137 stars), without indication of intrinsic 
star-to-star scatter.

From the forbidden [O {\sc i}] lines at 6300.3, 6363.8~\AA \ and the Na
doublets at 5682-88 and 6154-60~\AA\ we measured O and Na abundances for a
large sample of stars. The [Na/Fe] versus [O/Fe] ratios follow the well known
Na-O
anticorrelation, signature of proton-capture reactions at high temperature,
found in all other GCs examined so far.
The anticorrelation in NGC 6752 is the same at every luminosity along the RGB,
suggesting that any mixing effect must be negligible.

We also derived the distribution function of stars in
[O/Na], i.e. along the Na-O anticorrelation, finding it similar to the one
in M~13, a possible indication of a relation with HB morphology. However, the
degree of chemical anomalies is not as severe as that in M~13.
Further light on the subject is expected once we have completed the
analysis of our GC survey, covering the whole range of cluster physical
parameters.

\begin{acknowledgements}
We thank Dr. D. Yong for providing unpublished EWs.
This
publication makes use of data products from the Two Micron All Sky Survey,
which is a joint project of the University of Massachusetts and the Infrared
Processing and Analysis Center/California Institute of Technology, funded by
the National Aeronautics and Space Administration and the National Science
Foundation. This work was partially funded by the Italian MIUR
under PRIN 2003029437. We also acknowledge partial support from the grant
INAF 2005 ``Experimental nucleosynthesis in clean environments". 
\end{acknowledgements}


\begin{thebibliography}{}

\bibitem[{Alonso et al.} {1999}]{alonso99} 
 Alonso, A., Arribas, S. \& Martinez-Roger, C. 1999, A\&AS, 140, 261 
\bibitem[{Alonso et al.} {2001}]{alonso01} 
 Alonso, A., Arribas, S. \& Martinez-Roger, C. 2001, A\&A, 376, 1039 
\bibitem[]{} Asplund, M. 2005, ARA\&A, 43, 481
\bibitem[{Bragaglia et al.} {2001}]{bragaglia01} 
 Bragaglia, A., Carretta, E., Gratton, R.G. et al. 2001, AJ, 121, 327
\bibitem[]{} Brown, J. H., Burkert, A., Truran, James W. 1991, ApJ, 376, 115
\bibitem[]{} Brown, J. H., Burkert, A., Truran, James W. 1995, ApJ, 440, 666
\bibitem[]{} Buonanno, R., Caloi, V., Castellani, V., Corsi, C., 
 Fusi Pecci, F., Gratton, R. 1986, A\&AS, 66, 79 
\bibitem[]{} Buzzoni, A., Fusi Pecci, F., Buonanno, R., Corsi, C.E. 1983, A\&A,
128, 94
\bibitem[{Cardelli et al.} {1989}]{cardelli89} 
 Cardelli, J.A., Clayton, G.C., \& Mathis, J.S. 1989, ApJ, 345, 245
\bibitem[]{} Carretta, E. 1994, Ph.D. Thesis, University of Padova
\bibitem[]{} Carretta, E. 2006, AJ, 131, 1766
\bibitem[]{} Carretta, E., Gratton R.G. 1997, A\&AS, 121, 95 
\bibitem[]{} Carretta, E., Gratton R.G., Lucatello, S., Bragaglia, A., 
  Bonifacio, P. 2005, A\&A, 433, 597 
\bibitem[{Carretta et al.} {2006}]{carretta06} 
Carretta, E., Bragaglia, A., Gratton R.G., Leone, F., Recio-Blanco,
 A., Lucatello, S. 2006, A\&A, 450, 523 (Paper I)
\bibitem[]{} Cayrel, R. 1986, A\&A, 168, 8
\bibitem[]{} Cohen, J.G. \& Melendez, J. 2005, AJ, 129, 303
\bibitem[]{} D'Antona, F. \& Caloi, V. 2004, ApJ, 611, 871
\bibitem[]{} Denisenkov, P.A., Denisenkova, S.N. 1989, A.Tsir., 1538, 11
\bibitem[{Gratton et al.} {1999}]{gratton99} 
 Gratton, R.G., Carretta, E., Eriksson, K., \& Gustafsson, B. 1999,
 A\&A, 350, 955 
\bibitem[]{} Gratton, R.G., Sneden, C., Carretta, E., Bragaglia, A. 2000, A\&A,
 354, 169 
\bibitem[{Gratton et al.} {2001}]{gratton01} 
 Gratton, R.G., Bonifacio, P., Bragaglia, A., et al. 2001, A\&A, 369, 87
\bibitem[]{} Gratton, R.G., Sneden, C., \& Carretta, E. 2004, ARA\&A, 42, 385
\bibitem[{Gratton et al.} {2003a}]{gratton03a} 
 Gratton, R.G., Bragaglia, A., Carretta, E., Clementini, G.,
 Desidera, S., Grundahl, F., Lucatello, S. 2003a, A\&A, 408, 529
\bibitem[{Gratton et al.} {2003b}]{gratton03b} 
 Gratton, R.G., Carretta, E., Claudi, R., Lucatello, S., \&
 Barbieri, M. 2003b, A\&A, 404, 187
\bibitem[{Gratton et al.} {2005}]{gratton05} 
 Gratton, R.G., Bragaglia, A., Carretta, E., De Angeli, F.,
 Lucatello, S., Piotto, G., Recio Blanco, A. 2005, A\&A, 440, 901 
\bibitem[]{} Grundahl, F., Briley, M., Nissen, P.E., Feltzing, S. 2002, A\&A,
 385, L14
\bibitem[]{} Harris, W.~E. 1996, AJ, 112, 1487
\bibitem[]{} Kraft, R.P. 1994, PASP, 106, 553
\bibitem[{Kurucz}{1993}]{kur93} 
 Kurucz, R.L. 1993, CD-ROM 13, Smithsonian Astrophysical
\bibitem[]{} Langer, G.E., Hoffman, R., \& Sneden, C. 1993, PASP, 105, 301
\bibitem[]{} Magain, P. 1984, A\&A, 134, 189
\bibitem[{Momany et al.} {2004}]{momany04} 
Momany, Y., Bedin, L.R., Cassisi, S. et al. 2004, A\&A, 420, 605
\bibitem[]{} Moni Bidin, C., Moehler, S., Piotto, G., Recio Blanco, A.,
 Momany, Y., Mendez, R.A. 2006, A\&A, 451, 499
\bibitem[]{} Norris, J.E., \& Da Costa, G.S. 1995, ApJL, 441, L81
\bibitem[]{} Parmentier, G. \& Gilmore, G. 2001, A\&A, 378, 97
\bibitem[{Pasquini et al.} {2002}]{pasquini02} 
Pasquini, L. et al. 2002, The Messenger, 110, 1
\bibitem[]{} Prantzos, N., Charbonnel, C. 2006, A\&A, 458, 135 
\bibitem[{Pritzl et al.} {2005}]{pritzl05}
 Pritzl, B.J.. Venn, K.A.. Irwin, M. 2005, AJ, 130, 2140
\bibitem[]{} Ramirez, S. \& Cohen, J.G. 2003, AJ, 125, 224
\bibitem[{Rosenberg et al.} {1999}]{rosenberg99}
 Rosenberg A., Saviane I., Piotto G., Aparicio A., 1999, AJ, 118, 2306 
\bibitem[]{} Sandquist, E.L. 2000, MNRAS, 313, 571
\bibitem[]{} Skrutskie, M.F. et al. 2006, AJ, 131, 1163
\bibitem[]{} Sneden, C., Kraft, R.P., Guhathakurta, P., Peterson, R.C.,
  Fulbright, J.P. 2004, AJ, 127, 2162
\bibitem[]{} Ventura, P. D'Antona, F., Mazzitelli, I., \& Gratton, R. 2001,
  ApJ, 550, L65
\bibitem[]{yong03} 
 Yong, D., Grundahl, F., Lambert, D.L, Nissen, P.E., Shetrone, M.D. 2003,
 A\&A, 402, 985
\bibitem[{Yong et al.} {2005}]{yong05} 
 Yong, D., Grundahl, F.,  Nissen, P.E., Jensen, H.R., Lambert, D.L. 2005,
 A\&A, 438, 875
	

\end{thebibliography}
\end{document}